\documentstyle[epsfig,12pt]{article}
\begin{document}
\setlength{\baselineskip}{5mm}
\newcommand\be{\begin{equation}}
\newcommand\ee{\end{equation}}
\newcommand\bea{\begin{eqnarray}}
\newcommand\eea{\end{eqnarray}}
\newcommand\ket[1]{|#1\rangle}
\newcommand\bra[1]{\langle #1|}
\newcommand\braket[2]{\langle #1|#2\rangle}
\newcommand{\fatalpha}{{\bf \alpha \kern -0.44em \alpha}}
\newcommand{\fatsigma}{{\bf \sigma \kern -0.54em \sigma}}
\newcommand{\tpchi}{{\bf \chi \kern -0.35em \chi}}
\newcommand{\llambda}{{\bf \lambda \kern -0.45em \lambda}}
\newcommand\tr{\mbox{ Tr }}
\newcommand\ceq[2]{
 \begin{minipage}{#1}{
  \vspace{4pt}
   \begin{center}#2
   \end{center}
  \vspace{-10pt}}
 \end{minipage} }
\renewcommand{\thefootnote}{\fnsymbol{footnote}}
\newcommand \clebsch[6]{\left<\begin{array}{cc|c}
   #1 & #2 & #3 \\ #4 & #5 & #6 \end{array} \right>} 
\newcommand\wignerTroisJ[6]{\left(\begin{array}{ccc}
   #1 & #2 & #3 \\ #4 & #5 & #6 \end{array} \right)}
\newcommand\bsigma{\sigma\hspace{-.56em}\sigma}
\newcommand\brho{\rho\hspace{-.465em}\rho}
\def\Z{\mbox{Z\hspace{-.33em}Z}}              

\noindent{\large\bf
The Wigner Kernel of a Particle obtained from the Wigner Kernel of a Spin 
by Group Theoretical Contraction\footnote{Proceedings of the {\em XIII International Colloquium on Group Theoretical Methods in Physics},  31 July  - 5 August, 2000, Dubna, Russia} 
}\vspace{4mm}

\noindent{
J.-P. Amiet and \underline{St. Weigert} 
}\vspace{1mm}

\noindent{\small
Institut de Physique, Universit\'e de Neuch\^atel \\
Rue A.-L.\ Breguet 1, CH-2000 Neuch\^atel, Switzerland 

}\vspace{4mm}

\subsection*{Outline}
The Moyal formalism for a particle can be derived from the Moyal formalism for a spin. 
This is done by contracting the group of rotations to the oscillator group. 
A new derivation is given for the contraction of the spin Wigner-kernel to the
Wigner kernel of a particle. 

\subsection*{Introduction}

A {\em symbolic calculus} is a one-to-one correspondence between (self-adjoint) operators $\widehat A$ acting on a Hilbert space $\cal H$ of a quantum system, and (real) functions $W_A$ defined on the phase-space $\Gamma$ of the corresponding classical system (see [1] for a summary). Representating quantum mechanics in terms of $c$-number valued 
functions has various appealing properties since it allows one to  
situate the quantum mechanical description of a system in a familiar frame. 
The visualisation of quantum states and operators in classical phase space helps to develop an intuitive understanding of quantum features. Furthermore, it is interesting from a structural point of view: to calculate expectation values of operators by means of `quasi-probabilities' in phase space is strongly analogous to the determination of mean values in classical statistical mechanics [2].

The quantum mechanics of spin and particle systems can be
represented faithfully in terms of functions defined on the surface 
of a sphere with radius $s$, and on a plane, respectively. Intuitively, one 
expects these
phase space-formulations to approach each other for increasing values of the
spin quantum number since the surface of a sphere is then approximated by a plane with increasing accuracy. Therefore, appropriate Wigner functions of a spin, say, should go over smoothly into particle Wigner-functions in the limit of large $s$. Two different approaches [3,4] have confirmed this using the group theoretical technique of {\em contraction} [5] which map $U(2)$ to the oscillator group. 

In Ref. [3], the transition of the Wigner kernel of the spin to the Wigner kernel of a particle has been reduced to the evaluation of the limit of certain 
sums over Clebsch-Gordan coefficients. If these sums take specific values--and only those values--, the operator kernel, which  characterizes in condensed form the symbolic calculus of a spin, goes over smoothly to the corresponding particle kernel. The present contribution contains a new method to evaluate the 
sums in question which is, in fact, the difficult part of the transition from the spin to the particle formalism. In the following, the notation of Ref. 
[3] is employed, and the reader will find there the details on the underlying contraction procedure. Here the focus is on a technical problem, namely to sum a particular series. A brief summary at the end puts the result of the calculation into perspective. 
\subsection*{Summing the series}
Consider the the numbers 
\be \label{sum}
S_n= \lim_{s\to \infty} \, \, \sum_{l=0}^{2s} 
                 \left(\frac{2l+1}{2s+1}\right)^{1/2} 
                 \clebsch{s}{s}{l}{s-n}{n-s}{0} \, ,  \qquad n =0,1,2,\ldots
\ee
where each term of the sum is a multiple of a Clebsch-Gordan 
coefficent [6]. As shown in [3], the Wigner kernel of a spin turns into the Wigner 
kernel of the particle if 
\be \label{sumtoprove}
S_n = 2 \, , \qquad n=0,1,2,\ldots
\ee
holds. Therefore, (\ref{sumtoprove}) requires that there exist 
infinitely many $n$-independent sum rules for Clebsch-Gordan 
coefficients which are, apparently, not available in the literature. 
It is the purpose of this contribution to prove Eq. (\ref{sumtoprove}) 
in a way {\em different} from the one given in [3].  

The starting point is a recurrence relation satisfied 
by Clebsch-Gordan coefficients [6]:
\bea \label{CGrec}
 [ l(l+1) & -&2s(s+1) + 2m^2 ] \clebsch{s}{s}{l}{m}{-m}{0} \nonumber \\ 
&=& [s(s+1) - m(m+1)] \clebsch{s}{s}{l}{m+1}{-(m+1)}{0} \nonumber \\
&   &  +\,  [s(s+1) - m(m-1)] \clebsch{s}{s}{l}{m-1}{-(m-1)}{0}  \, .
\eea
Define the quantities
\be \label{capitalD}
D_{k,n}^s = \sum_{l=0}^{2s} 
                      \left(\frac{l(l+1)}{2s+1}\right)^k
                   \left(\frac{2l+1}{2s+1}\right)^{1/2}
                             \clebsch{s}{s}{l}{s-n}{n-s}{0} \, .
\ee
Multiply the recurrence (\ref{CGrec}) by 
$((2l+1)/(2s+1))^{1/2}(l(l+1)/(2s+1))^{k}$ and sum over $l=0,1,2,\ldots, 2s$, which implies that 
\bea \label{D0}
 D_{k+1,n}^s & = & \left(1-\frac{n+1}{2s+1}\right)(n+1) D_{k,n+1}^s + 
\left(1-\frac{2n^2+2n+1}{(2s+1)(2n+1)}\right)(2n+1) D_{k,n}^s \nonumber \\
    &   &          +\left(1-\frac{n}{2s+1} \right)n D_{k,n-1}^s
\eea
with $n$ taking any integer value from $0$ to $2s$. 
Taking the limit $s\to \infty $ with fixed $n$ the coefficients in large brackets become equal to unity and one obtains 
\be \label{limitDK}
D_{k+1,n} = (n+1) D_{k,n+1} + (2n+1) D_{k,n} +n D_{k,n-1}   \, ,
\ee
where $D_{k,n}$ is defined as the limiting value of $D_{k,n}^s$,
\be \label{defineD}
D_{k,n} = \lim_{s\to \infty} D_{k,n}^s \, , \qquad k= 0,1,2,\ldots
\ee
Comparison with (\ref{sum}) shows that $ S_n = D_{0,n}$. These numbers can be calculated in the following way. First, one shows that $D_{k,0}$ and $D_{0,n}$ are related by 
\be \label{sumrule}
D_{k,0} = k! \sum_{n=0}^k {k \choose n} D_{0,n} \, ;
\ee
second, one calculates explicitly the value of $D_{k,0}$ which is found to be 
\be \label{initial}
D_{k,0} = 2^{k+1} k! \, .
\ee
These two identities will be derived in the following section.
Combining them leads to 
\be 
\sum_{n=0}^k {k \choose n} D_{0,n} = 2^{k+1} \, ,
\ee
valid for each $k=0,1,2,\ldots$ It is straightforward now to determine 
from $k=0$ that $D_{0,0} = 2$. Induction on $k$ implies that $D_{0,n} = 2$ for all $n$, and the final result reads 
\be
S_n  = 2\, , \qquad n=0, 1, 2, \ldots 
\ee 
\subsection*{Two identities}
In order to prove Eq. (\ref{initial}) write down the expression for $D_{k,0}$ 
according to (\ref{capitalD}) in the limit of 
large values of $s$, 
\be \label{approxsum}
D_{k,0} = \lim_{s\to \infty} \, \, \sum_{l=0}^{2s} 
               \left(\frac{l(l+1)}{2s+1}\right)^k \left(\frac{2l+1}{2s+1}\right)^{1/2}
                             \clebsch{s}{s}{l}{s}{-s}{0} \, .
\ee
One can simplify this expression by approximating 
the Clebsch-Gordan coefficients 
\bea \label{lowestdelta}
\left(\frac{2l+1}{2s+1}\right)^{-1/2} \clebsch{s}{s}{l}{s}{-s}{0} 
 & \equiv & \left( \frac{(2s)!}{(2s-l)!} \frac{(2s)!}{(2s+l+1)!}\right)^{1/2}
\nonumber \\
&\equiv&  \left( \frac{\Pi_{k=0}^l (1- k/(2s+1) )}{\Pi_{k=0}^l 
(1+ k/(2s+1) )} \right)^{1/2} \nonumber \\ 
&\sim& \exp \left[- \frac{1}{2} \frac{l(l+1)}{2s+1} \right] \, ,
\eea
where has been used the approximation
\be \label{lowestterm}
\left( 1 - \frac{k}{2s+1}\right) \sim \exp [ -\frac{k}{2s+1} ] \, , 
\ee
valid for each finite $k$ and large values of $s$.  Upon introducing  
\bea
x_l &=& \frac{1}{2} \frac{l(l+1)}{2s+1} \\
\Delta x_l &=& (x_{l+1} - x_l) = \frac{1}{2} \frac{2l+1}{2s+1}+ {\cal O} (1/s) \, ,
\eea
the expression (\ref{approxsum}) is seen to be a Riemann sum 
defining an integral which is easily evaluated,
\be
D_{k,0} = 2^{k+1} \int_0^\infty dx \, x^k e^{-x} 
            = 2^{k+1} k! \, ,
\ee
confirming Eq. (\ref{initial}).

Let us turn to the sum rule stated in Eq. (\ref{sumrule}). Consider 
the quantities
\be \label{generalsumrule}
T_k^N = N! \sum_{n=0}^N {N \choose n} D_{k-N,n} \, , \qquad N=0,1,2, \ldots , k \, .
\ee
For a fixed value of k, the value of the sum on the right-hand-side is {\em independent} of the value of $N$,
\be \label{indep}
T_k^N = T_k^{N'} \, , \qquad N,N' =  0,1,2, \ldots , k \, .
\ee
This follows from a straightforward calculation exploiting the recurrence relation (\ref{limitDK}): 
\bea \label{firstSR}
T_k^{N-1} &=& 
N! \sum_{n=0}^{N-1} {N-1 \choose n} D_{k-N+1,n}  \\
&=& (N-1)! \sum_{n=0}^{N-1} {N-1 \choose n} 
        \left[ (n+1) D_{k-N,n+1}  + (2n+1) D_{k-N,n} + n D_{k-N,n-1} \right] \nonumber \\
&=&  (N-1)! \sum_{n=0}^{N-1} 
          \left[ {N-1 \choose n-1} n 
                   + {N-1 \choose n} (2n+1) 
                   + {N-1 \choose n+1}  (n+1) \right] D_{k-N,n} \nonumber
\eea
where the last identity is due to appropriately relabeling the summation index. Evaluating the expression in square brackets gives 
\be \label{bracket}
N \, \frac{N!}{n! (N-n)!} \qquad \mbox{or equivalently}\qquad \frac{1}{(N-1)!} \, N! \, {N \choose n} \, ,
\ee
which implies Eq. (\ref{indep}),
\be \label{secondSR}
T_k^{N-1}  = N! \sum_{n=0}^N {N \choose n} D_{k-N,n} 
                \equiv T_k^N \, . 
\ee
Setting now $N=0$ and $N' = k$ in (\ref{indep}), one obtains $T_k^0 = T_k^k$, or, explicitly, 
\be \label{sumrule2}
D_{k,0} = k! \sum_{n=0}^k {k \choose n} D_{0,n} \, ,
\ee
which is the identity (\ref{sumrule}) required for the proof given in the previous section.
\subsection*{Discussion}
The calculation presented here provides to an elementary proof that the kernel defining the familiar Wigner formalism for a spin becomes, in the limit of infinite values of $s$, the Wigner kernel of a particle. As the kernel defines 
entirely a phase-space representation, this result guarantees that the Moyal formalism for a particle is reproduced automatically and {\em in toto}, if the limit $s\to \infty$ of the spin Moyal formalism is taken. 

This result shows that contraction of groups is a useful tool in order to establish structural analogies between different phase-space representations {\em \`a la} Wigner. It is expected that similar relations can be found among other phase-space representations of quantum systems possessing Lie-group symmetries [1].   
\subsection*{Acknowledgements}
St. W. acknowledges financial support by the {\em Schweizerische Nationalfonds}.


\newcommand{\etal}{{\em et al.}}
\setlength{\parindent}{0mm}
\vspace{5mm}
{\bf References}
\begin{list}{}{\setlength{\topsep}{0mm}\setlength{\itemsep}{0mm}%
\setlength{\parsep}{0mm}}
%
\item[1.] C.\ Brif and A.\ Mann,  Phys.\ Rev.\ A {\bf 59}, 971 (1999).
\item[2.] E.\ P. Wigner, Phys. Rev. {\bf 40}, 749 (1932).
\item[3.] J.-P.\ Amiet and St.\ Weigert, Phys.\ Rev.\ A, (in press).
\item[4.] A.\ B.\  Klimov and S.\ M.\ Chumakov,  J.\ Opt.\ Soc.\ Am.\ (in press). 
\item[5.] F.\ T.\ Arecchi, E.\ Courtens, R.\ Gilmore, and H.\ Thomas,
Phys.\ Rev.\ A {\bf 6} 2211 (1972).
\item[6] N.\ Ja.\ Vilenkin: {\em Fonctions sp\'eciales et Th\'eorie de la Repr\'esentation des Groupes},
(Dunod, Paris 1969).

%
\end{list}

\end{document}